# AN EFFICIENT SECURITY MECHANISM FOR HIGH-INTEGRITY WIRELESS SENSOR NETWORKS

Jaydip Sen
Convergence Innovation Lab, Tata Consultancy Services
Bengal Intelligence Park, M2 & N2, Sector- V, Block- GP, Salt Lake Electronics Complex,
Kolkata-700091, INDIA.
Email: jaydip.sen@tcs.com

**ABSTRACT**
*Wireless sensor networks (WSNs) have recently attracted a lot of interest in the research community due their wide range of applications. Unfortunately, these networks are vulnerable to numerous security threats that can adversely affect their proper functioning. This problem is more critical if the network is deployed for some mission-critical applications such as in a tactical battlefield. Random failure of nodes and intentional compromise of nodes by an insider attack in a WSN pose particularly difficult challenges to security engineers as these attacks cannot be defended by traditional cryptography-based mechanisms. In this paper, a security solution is proposed for detecting compromised and faulty nodes in a WSN. The mechanism also isolates a compromised node from the network so that it cannot participate in any network activity. The proposed mechanism is based on misbehavior classification, behaviour monitoring and trust management. It involves minimum computation and communication overhead and is ideally suited for a resource-constrained, high-integrity WSN.*

**KEY WORDS**
Security, insider attack, trust management, misbehavior classification, reputation.

## 1. Introduction

Wireless sensor networks (WSNs) consist of hundreds or even thousands of small devices each with sensing, processing, and communication capabilities to monitor the real-world environment. They are envisioned to play an important role in a wide variety of areas ranging from critical military surveillance applications to forest fire monitoring and the building security monitoring in the near future. In these networks, a large number of sensor nodes are deployed to monitor a vast field, where the operational conditions are most often harsh or even hostile. To operate in such environments, these networks should be equipped with security mechanisms to defend against attacks such as node capture, physical tampering, eavesdropping, denial of service, etc [1, 27, 28]. Security mechanisms deployed in these networks should involve collaborations among the nodes due to the decentralized nature and absence of infrastructure. Moreover, due to the large varieties of applications and diverse environmental conditions in which WSNs operate, it is necessary that security mechanisms are designed in context to the application domain. Therefore, the design of a secure WSN should consider the application-specific security context. Unfortunately, most of the currently existing WSN security solutions consider an abstract network model based on assumptions that are not coupled with the application specific details. As a result, many of these assumed vulnerabilities are theoretical in nature without any consideration to application-specific threats and requirements.

Apart form the lack of application specific objectives, the current security mechanisms for WSNs have other problems too. Although, some of the existing security solutions for defense against outsider attacks by key management schemes [2, 3, 6, 24, 25] and secure node-to-node communication mechanisms [24, 26] are quite effective, these mechanisms break down even if a single legitimate node in the network is compromised [18]. Presence of a few compromised nodes can pose severe security threats in a WSN, as these nodes can launch different types of attacks e.g., dropping of legitimate reports, injection of bogus sensing reports, advertisement of inconsistent routing information, eavesdropping on in-network communication using exposed keys, etc. Such disruption by an insider attack can be very detrimental to the overall objective and mission of the WSN. Thus security counter-measures against each type of such attacks is very much in demand.

In the past few years, many cryptography-based security designs have been proposed for WSNs. Cryptography provides a number of efficient mechanisms for implementing data confidentiality, data integrity, node authentication, secure routing and access control in different types of networks. While these techniques are very useful in building secure WSNs, they are not sufficient to address the security threats posed by compromised and faulty sensors [18]. This is because any compromised or faulty node possesses the cryptographic key by virtue of being a legitimate member of the network, and thus, it is allowed to participate in the network activities in the same way as other nodes in the network. As cryptographic mechanisms are not sufficient to address threats posed by compromised and faulty sensor nodes, a more sophisticated technique is needed to fully address the problem.

Some propositions already exist in the literature for addressing the threats posed by compromised and faulty nodes in WSNs [19, 20, 21, 22, 23]. These mechanisms mainly focus on utilization of the redundancy in a WSN to improve its resilience against attacks. These techniques generally involve introduction of a threshold property in their designs to have robustness against a certain threshold (maximum) number of compromised and faulty nodes. However, in practice, these passive approaches are not very effective since the designed threshold parameter may deviate significantly from the real-world scenario.

To systematically address the problem of insider attack, this paper presents a unified proactive approach towards design of a defense mechanism against different types of attacks in a WSN. The proposed framework is based on a proactive data security mechanism that consists of two broad modules: (i) misbehavior characterization and monitoring, and (ii) trust management. While the first module categorizes different types of misbehavior of nodes and defines a set of monitoring criteria for each of these misbehaviors, the second module develops a trust management framework that evaluates the detection results of the first module and updates a *reputation table* in each sensor node. The communication among the sensors is based on their reputation records. The nodes having reputation values lower than a threshold are not allowed to participate in network activities.

The rest of the paper is organized as follows. Section 2 presents some related work on cryptography-based security mechanisms for WSNs. In Section 3, different security vulnerabilities of WSNs are discussed and the limitations of the currently existing mechanisms are identified. The proposed security scheme is described in Section 4 and Section 5. Section 4 presents misbehavior classification and handling module of the scheme. Section 5 describes the design principles of the trust management component. Section 6 presents some experimental results and the plan for further experiments. Section 7 concludes the paper.

## 2. Related Work

Most of the currently existing security schemes for WSNs are based on cryptographic mechanisms. In this section, some of these schemes are described briefly.

Eshenauer and Gligor [2] have proposed a random key pre-distribution scheme to establish keys between sensor nodes. In this scheme, each sensor node of a WSN receives a random subset of keys from a key pool before the deployment of the network. Any pair of nodes which are able to find one common key within their respective key subsets can use that key as a shared secret to initiate communication between them.

Chan, Perrig and Song [3] have proposed three models for pairwise key pre-distribution in a WSN. The first model is based on a $q$- composite scheme that extends Eschenaeuer's scheme by imposing an additional constraint. This constraint requires that for any two nodes to form a secure channel between them, they must have $q$ common keys between them. The second model involves a multi-path key reinforcement scheme that offers increased security against an adversary by performing key updates across multiple paths. The third model features a *random-pairwise key scheme* which ensures security of the network even in presence of a certain threshold number of compromised nodes.

Blom [4] has proposed a key pre-distribution scheme that enables any pair of nodes in the network to generate a secret pairwise key between them. Du and Deng [29] have described a scheme that offers improved network resilience by combining Blom's scheme with the scheme proposed by Eschenauer and Gligor.

Blundo et al [5] have described some schemes of key management for WSNs. These schemes allow any group of $n$ network nodes to compute a common key. This key is perfectly secret and secure against any collusion attempt by any t other nodes in the networks.

Liu and Ning [6] have presented a framework for pairwise key establishment based on polynomial key pre-distribution protocol proposed by Blundo et al [5] and the probabilistic key distribution of Chan, Perrig and Song [3]. They have further developed two novel pairwise key pre-distribution schemes- (i) random subset assignment and (ii) grid-based key pre-distribution scheme.

Law, Etalle, and Hartel [7] have investigated the issues related to group-wise pre-deployed keying and secret sharing of pre-deployed keying. They have proposed a group-wise scheme called '$k$-secure $t$-limited *group-*wise pre-deployed keying' where $k$ denotes the number of attacker groups and $t$ denotes the maximum number of communication groups. As this scheme is based on Blundo's key distribution scheme, it is secure from information theoretic point of view. The authors have also proposed a secret sharing scheme of pre-deployed keys known as *private-key sharing and pre-deployed keying*.

## 3. Limitations of the Current Schemes

The associated threats and vulnerabilities of a WSN are dependent on its application. As a consequence, WSN security design and analysis must be sensitive to its application context. Otherwise, the assumptions made on the organization of the WSN and the corresponding threats may become inconsistent with the problem domain. This will obviously lead to design of mechanisms that attempt to address unrealistic problems. Unfortunately, most of the current security solutions for WSNs suffer from this drawback. The design of a security scheme for a WSN should incorporate the security context that is not merely a precise technical specification; rather it is a set of security-related factors that narrow down the WSN design space to a region consistent with its application and environmental contexts.

Apart from the above generic limitations, there are other shortcomings with the current security schemes. Most of these schemes are based on cryptographic techniques, and therefore, these schemes are not capable of ensuring data security in WSNs. We will now illustrate this with the help of an example. Let us consider a tactical battlefield scenario where a WSN is deployed to provide dynamic location information of the adversary soldiers to the legitimate users (soldiers) of the network. We assume that some of the sensors in the network are compromised and these compromised nodes are under the complete control of the adversary soldiers. The cryptographic mechanisms are supposed to be already deployed in the WSN. We now illustrate three attacks that cannot be addressed by cryptographic primitives.

*(i) Attacks on data confidentiality:* For confidentiality of sensitive data, it is required that no adversary should be able to access network data even if he is able to compromise some nodes in the network. However, in the tactical battlefield scenario, a compromised node may initiate a query asking for the location of a legitimate soldier and send this information to an adversary soldier. Since the query sent by the compromised node is authenticated with the cryptographic key, cryptographic mechanisms cannot prevent this attack on data confidentiality.

*(ii) Attacks on data authenticity:* The network should allow only authenticated nodes to participate in the network activities. Thus only authentic data that covey the actual status of the environment should be communicated and processed by the network nodes. Spurious data from compromised and faulty nodes should be either prohibited from being injected in the network or filtered out after they are received by the network nodes. However, in the example scenario, a compromised or faulty node is freely allowed to send false information in reply to a query of a legitimate soldier, thereby completely misleading the latter. Cryptographic mechanisms cannot prevent this attack as the compromised nodes possess the cryptographic key and these nodes are authenticated members of the network. Thus messages sent by these nodes are also assumed to be authentic messages. A few attack-resilient approaches have been proposed to mitigate this attack. These mechanisms usually introduce a threshold property in their designs and thus gain robustness against a certain maximum number of compromised and faulty nodes. For instance, in order to prevent compromised nodes from reporting false alarm to a sink node, observations from multiple sensors (say $t$) may be utilized. As long as there are not more than $t$ compromised nodes in the network, the event report will be correct. However, the effectiveness of these approaches is doubtful in practical situations where the predefined threshold may be significantly different from the real scenario in the network.

*(iii) Attacks on data availability:* Critical information in the network should be always available on demand to any legitimate user (soldier). On the other hand, an adversary should not be allowed to access any critical information in the network. An attack on data availability will try to make critical data not available to the legitimate users in the network. In the example scenario, a compromised node may intentionally drop an alarm message sent by a legitimate node in the network without forwarding it to its neighbors. This is an attack on data availability as the legitimate nodes in the network will not receive the alert message resulting in a disaster. It is quite clear that cryptographic mechanisms are not capable to defend against this attack too.

Above examples illustrate that cryptography-based techniques are not sufficient to address security threats posed by compromised and faulty nodes in WSNs. A more proactive approach is needed where compromised and faulty nodes are promptly identified and prevented from participating in network activities.

### 3.1 WSN vulnerabilities and their impact on design

There are many characteristic features of WSNs that make these networks particularly vulnerable to different types of attacks. From an attacker's perspective, the opportunity for a particular type of attack is essentially the ease or difficulty in launching that attack. The opportunity, when combined with the benefit, can be used to define cost-benefit ratios for different types of attacks. Because of their data-driven nature and large number of possible threats, WSNs are vulnerable to different types of attacks. Some of them are as follows:

*(i) Physical attack on the sensors:* The in-situ nature of WSNs requires that sensors be integrated with the environment they are monitoring. As a result, the network may be physically vulnerable depending on the nature and extent of the sensor field. Access to the sensors can be used to physically destroy them, or to capture and subvert them to collect confidential data or to make an attempt in launching insider attack on the network.

*(ii) Attack on the wireless channel:* In addition to physical vulnerability of sensors, attackers may have access to anything transmitted over the wireless channel due to broadcast nature of the channel. Further, attackers can launch an outsider attack by injecting their packets to cause interference with legitimate packet transmission in the network.

*(iii) Attack on coordination and self-configuration:* For proper functioning and operation, the nodes in a WSN require coordination among them and self-configuration of the network via distributed protocols with localized interactions [8]. In many applications, WSNs heavily rely on coordinated services such as routing, localization, time synchronization, and in-network data processing for self-configuration and collaborative processing of data. Unfortunately, these services present unique opportunities of attacks which are absent in conventional networks. For example, compromised nodes can claim a false proximity to the sink node to attract packets from other nodes, considerably increase the clock skew to disrupt coordinated network operations such as sleep scheduling, and inject false data to reduce the accuracy of sensing. These attacks on fundamental coordination and self-configuration functions can be detrimental to the sensor network.

*(iv) Attacks due to visibility of the network:* A good understanding of the span and structure of the network opens up risks for more precise and effective attacks. The information about the network such as expected mission lifetime, deployment of the nodes, communication among the nodes and access to the sensor field can provide crucial data to a potential attacker. Beyond mere detection of the presence of a network, detection of its structure, topology and other communication pattern among the nodes can invite more directed attacks, e.g., targeted attack on the base station or nodes which are in the vicinity of the attacker. A good range of design choices for WSNs are available in terms of the types of sensors and their capabilities, sensor density and distribution, and application software. Typically, the design of these elements is driven by cost, energy-efficiency and application-level performance such as the coverage or accuracy.

In applications with high attacker motivation, the WSN design should make a trade off between cost and network performance to keep vulnerabilities at acceptable levels. At the physical level, this translates into using more expensive and secure sensors, or deploying more sensors to introduce redundancy or tolerance against a possible attack. For example, the network can be better protected by deploying multiple base stations when the attacker's motivation for attacking a single base station is expected to be high. These extra capabilities may be introduced or tasked non-uniformly depending on the application; for example, more expensive and secure sensors may be delegated with critical roles in underlying services or may be used in less secure areas of the network.

In terms of protocols, services, and application software, the tradeoff between security and performance is more explicit. Vulnerabilities arise especially in the setup of critical services such as routing. Protocols such as geographic routing expose the location of the destination in each packet, which could in turn, invite attacks on critical points of the infrastructure. The use of encryption can improve confidentiality at the cost of energy and computational resources; the size of the encryption key makes this a tunable tradeoff. In addition, to protect the structure of the network, anomaly and intrusion detection as well as trust management approaches should be employed. They enable detection of attacks and tolerating them, if possible, by isolating misbehaving nodes. Using per-hop encryption facilitates in-network data processing but may make the network vulnerable to attacks launched by a group of compromised nodes. In critical applications, end-to-end encryption may be used, to achieve higher level security. However, it will not be energy-efficient as in-network processing will not be possible with end-to-end encryption in place.

## 4. The Proposed Security Mechanism

In this section, we describe the proposed security mechanism for a WSN. First we present the network model and the security model where we list down different assumptions about the networks and its applications. Then we discuss different types of misbehavior exhibited by a faulty or compromised sensor node. Finally, some rules are presented for handling different types of compromised nodes.

### 4.1 Network model and system security model

*Network Model:* We consider a WSN with a large number of uniformly distributed static sensors that monitors a vast terrain. The WSN may be deployed by techniques such as aerial scattering. Following assumptions are made about the network: (i) after deployment, each node has a localization mechanism by which it can get an idea about its authentic geographic location. (ii) the WSN is densely connected so that it can support fine-grained collaborative sensing and computing even in the event of node and link failures, (iii) multiple users may put query for network information simultaneously into the network, (iv) the deployed sensor nodes are not tamper-resistant and have limited communication range, (v) wireless links in the network are symmetric, (vi) for detection of any event or to resolve a user query, a collaborative participation of multiple sensors is possible.

*Security Model:* From security perspective, following assumptions about the network are made: (i) cryptography-based mechanisms are already deployed in the network. Thus, every sensor can authenticate itself to its neighbors by the cryptographic key it possesses, (ii) at the time of network bootstrapping, there are no compromised or faulty nodes in the network, (iii) during network operation an adversary can physically compromise a few number of sensors and gain full control on them. It is not possible for any legitimate sensor nodes in the network to detect and understand the communication of messages between the compromised nodes and the adversary.

In the proposed security framework, we assume that the nodes may fail due to a number of events such as radio failure, sensing function error, system crash etc. Since failed nodes also lead to generation of wrong (bogus) data, their presence is equally detrimental to the network functioning. Moreover, Byzantine failures of nodes may result in persistent, transient or probabilistic failure pattern which will be impossible to detect by a simple deterministic mechanism.

Finally, the adversary is assumed to be sophisticated and is driven by two main objectives:

(i) *Benefit form data:* The adversary wants to gain access to the sensitive data being monitored or relayed in the network. Thus, the goal of the attacker is to have an access to the data being carried.

(ii) *Mission interference:* The adversary wants to interfere with the mission of the WSN. In this case, the data inside the WSN is not necessarily of interest to the attacker. Instead, he wants to compromise the network's ability to function by injecting spurious data or by disrupting a set of nodes in the network. Disrupting the availability of the network is also one of the objectives of the adversary.

### 4.2 Types of misbehavior of nodes

The nodes in the network may exhibit different types of misbehavior due to insider attack or random failure. The insider attacks can be classified into four broad types: (i) data forwarding related, (ii) data generation related, (iii) routing related, and (iv) miscellaneous. Different types of data forwarding attack are: message delay attack, selective forwarding attack, message alteration attack, message replay attack, sinkhole attack, message collision attack etc. Data generation related attack includes spurious data injection attack, bogus query attack, report disruption attack etc. Routing related attack involves: hello flood attack, wormhole attack, spurious routing information attack, Sybil attack etc. Finally, Byzantine attack, node replication attack, node relocation attack etc. fall under miscellaneous types of insider attack. In Byzantine attack, a malicious node intermittently acts in proper way.

After having identified different types of insider attack in a WSN, we make the following general observations:
- Networks having cryptographic mechanisms already in place will be able to detect compromised nodes as these nodes will fail in authentication. These nodes will not be able to process and forward the packets arriving at them as they do not have the key, resulting in packet dropping. This packet dropping will be observed by the neighbour nodes and subsequently the compromised nodes will be identified.
- In case of data forwarding related attack, it will be more difficult to detect compromised nodes that selectively or randomly drop packets. The compromised nodes may act in collusion and drop packets multiple hops away, making a localized detection algorithm completely ineffective. This type of attack will not be possible, however, if the application requires end-to-end acknowledgment.
- A compromised node may occasionally report spurious sensed result to cheat its neighbors in a collaborative aggregation problem. However, the sensing results within a neighbourhood should not vary significantly and the nodes with significantly different sensing results should be suspected.
- Generally, it is very difficult to detect compromised nodes that launch message collision attack. However, a sensor does have the ability to detect abnormally high rate of collision by comparing its packet delivery ratio with its neighbors and their location information.
- A Byzantine attack compromises software platform of a sensor node by running some malicious code. Such an attack can be detected by using code attestation techniques implemented in the nodes [9].

In addition to the compromised nodes, the network may also have faulty nodes due to node failure. Failure of nodes may lead to one or more of the following events: (i) random message alteration, (ii) random message broadcast, (iii) sensing function error, and (iv) random packet dropping. However, these events pose the same security threats as the compromised nodes.

**4.3 Rules for handling different misbehavior**

Any attack on a WSN manifests itself in terms of occurrence of certain events. In this section, we formulate different rules for the proposed security schemes and also identify the events that will trigger application of these rules.

*Message acknowledgment rule:* In the proposed scheme, any unacknowledged message is treated as an evidence of an attack or a failure of the next-hop neighbor node, although the message packet might have been dropped by a sensor that is multiple hops away from the source node. To effectively handle the uncertainty associated with this, three possible cases are identified and dealt with in different manner. These cases are: (i) the acknowledgment successfully reaches the source node, (ii) the next-hop node of the source node has really forwarded the message but the acknowledgment has not reached the source node before the expiry of the timer, (iii) the next-hop has not forwarded the message and thus the acknowledgment would never reach the source.

*Authentication failure rule:* An authentication failure is considered as an evidence of an attack or node failure. The proposed scheme distinguishes between end-to-end and hop-wise authentication failures. Only hop-wise authentication failure is taken as an evidence of attack.

*Data validation rule:* It is assumed that the sensing results of a set of sensors in the same neighborhood follow a normal distribution. Thus if a node reports result that significantly deviates from the results of its neighbor nodes, it may be suspected to have been compromised or failed. The result consistency check is usually application specific, and the data abnormality may be detected given the context of the application.

*Traffic awareness rule:* The underlying MAC protocol in the WSN is assumed to be of 'collision avoidance' type. It is also assumed that the packet generation/relay actions of a sensor node can be sensed and estimated by its neighbor nodes. Any unexpected packet generation/relay by a node is thus considered as a sign of an attack or node failure.

*Packet delivery rule:* The packet delivery ratio of a node is defined as the ratio of the number of packets that are successfully delivered to a destination node to the number of packets that have been sent by the sender. In the proposed scheme, if a sensor node finds that its packet delivery ratio is below a threshold value, it treats this as a sign of message collision attack or a possible node failure and sends an alarm message to its neighbors. The threshold value of the packet delivery ratio is determined based on the quality of wireless links, and the probability of packet loss due to normal collision in the network. In contrast to all the previous rules, this is a self-evaluation rule where a sensor node evaluates itself rather than evaluating its neighbors.

*Memory consistency rule:* The memory status of a sensor node should be consistent and should maintain integrity. Any abnormal change in code size is a sign of an attack or a hardware failure.

*In-situ rule:* The proposed mechanism assumes a static WSN where every node keeps its location constant after the network deployment. Therefore, any location change of a node is a sign of a possible attack.

**4.4 Behavior monitoring**

Each sensor node is assumed to be operating in promiscuous mode and monitoring the packet forwarding activities of each of its neighbour nodes. For this purpose, each node maintains a buffer and randomly copies packets to it and checks whether these packets are forwarded by its neighbor nodes or dropped. Specifically the following activities are monitored by each node:

*Packet forwarding behaviour:* In the proposed mechanism, an algorithm is invoked periodically that senses the channel and compares ongoing data traffic with the recorded routing and MAC messages to detect any possible anomaly. In addition to this, timer and explicit acknowledgment mechanisms are used to detect any possible packet drop and packet duplication attempt.

*Time-space-data consistency:* In the proposed mechanism, the data validation information is obtained from the application modules of the neighbor nodes of a sensor. As the notion of data validation is application-specific, different data validation algorithms are designed for different applications.

*Traffic-related behavior:* It is assumed that the behavior of each neighbor nodes of a sensor with respect to message generation, relay, and duty cycle (sleep schedule) is possible to be estimated clearly from the context of the application. This is a realistic assumption in view of different statistical estimation techniques available for this purpose.

*Cryptographic failures:* The mechanism has the provision of capturing and storing every event of failure raised by the underlying cryptographic module. These events are signs of attacks or node failures in the network.

*Self-status:* Every node keeps track of its own packet delivery ratio, its memory status in real time, and its location information by some localization technique to identify any possible attack on it by an adversary.

This detailed set of behavior monitoring criteria provides the proposed security mechanism the capability to effectively detect any insider attack and presence of any failed nodes in the network. The design goal behind the formulation of such criteria is to keep monitoring activity of each individual sensor independent of each other. Each sensor node thus monitors its own neighbourhood and makes its own decision independent of the observation being made by its neighbors. This local detection algorithm avoids complexities of collaborative monitoring mechanisms and involves much less computation. However, due to limited local information available in the nodes, this local monitoring scheme will lead to some false positives- situations where the mechanism will raise alarm but actually there is no real attack or node failure. On the other hand, due to its simplicity, it will consume less power in the sensor nodes and will have less communication overhead compared to a collaborative monitoring mechanism. Moreover, the proposed mechanism follows a much simpler decision fusion and aggregation approach that is quite efficient and accurate in terms of attack detection capability. This is described in detail in the next section.

## 5. Trust Management Framework

As mentioned in Section 1, the proposed security mechanism has two broad modules- (i) misbehavior characterization and behavior modeling, (ii) trust management. The first module has been described in detail in Section 4. In this section, we present a description of the trust management module. Figure 1 describes the high-level architecture of the trust management module. The watchdog component enables each node to operate in promiscuous mode and collect first-hand information about the activities of each of its neighbors. The reputation and trust value of each node is computed on the basis of these observations. Based on these computed trust values, a sensor node's strategy for other node is determined. If the trust value is above a pre-determined threshold, then the strategy is to cooperate with the node otherwise not. If need arises, the trust management module of a node generates alert messages and broadcasts them in the network so that all the legitimate nodes in the network become aware about the presence of any compromised node in the network.

In computation of the reputation and trust metrics, higher weight is given to recent observations. Also a node with a higher reputation gets higher weight in its vote when a cooperative detection algorithm is invoked as discussed later in this section.

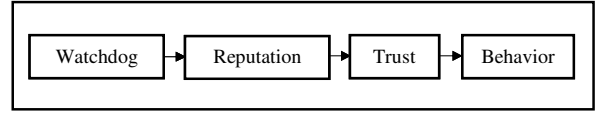

Figure 1: Architecture of the trust management module

### 5.1 Design principles of trust framework

The proposed trust management scheme is based on computation of trust and reputation metrics for the nodes in a WSN. The application of trust and reputation-based systems is quite common in peer-to-peer computing and mobile ad hoc networks. An extensive research has been done on these areas over the period of last five years [10, 11, 12, 13, 14]. In the following, we present the major design policies of the trust management mechanism of the proposed scheme.

*Policy1: During the network bootstrapping and initialization phase, each sensor node has the highest level of trust. Therefore, each node can fully trust all of its neighbors till it encounters experiences through interaction in the network which may reduce the initial trust value depending on the experience.*

The rationale for the assumption of all the nodes being equally trustworthy is due to that fact that these nodes are all deployed and scattered by the same trusted entity. Therefore, for a certain period of time at the initial period, all the nodes can be assumed to be good before some of them become faulty or compromised.

*Policy 2: The trust system is based on both good and bad experiences encountered. Thus, both positive and negative experiences are taken into account for computation of trust and reputation metrics of a node. These experiences are based on direct observations only.*

It is essential that both positive and negative experiences are taken into account for computation of a combined trust metric. Use of only positive information may lead to *false-praise attack* by a group of colluding sensors and thereby increasing their presence in terms of more number of compromised nodes. On the other hand, use of only negative information will lead to potential *bad-mouthing attack* where a group of colluding malicious nodes may send wrong information about a really good node thereby reducing its trust value and ultimately isolating it from the network. Thus, in the proposed scheme, both types of information are used for computation of reputation metric for a node. Moreover, to reduce the complexity of reputation computation, each sensor node maintains the reputation and trust information for its neighbor nodes only, and establishes a localized store of trust and reputation information.

*Policy 3: When the reputation value of a sensor node drops below a pre-determined threshold, the reputation fading mechanism is disabled.*

When the reputation value of a sensor node falls below a pre-determined threshold (a design parameter) in one or more of its neighbors, a voting algorithm is invoked (Policy 4). If the suspected node is identified to be really compromised or faulty, it is isolated and not allowed to participate further in any network activities. In some reputation systems for ad hoc networks, reputation fading mechanism is in place [12, 13, 15]. In such cases, a node that is isolated from the network due to low trust value is allowed after a certain period, when its reputation value is restored to a value higher than the minimum threshold due to reputation redemption. However, the proposed trust mechanism for WSNs does not employ reputation fading as sensor random failures are less frequent events. Thus, once the reputation value of a node falls below a threshold, it is highly likely that the node is compromised or has failed in hardware. In either case, the node should not be allowed to join the network further as there is no sensor recovery mechanism in place.

*Policy 4: A suspected sensor node is isolated from the network based on the outcome of a voting algorithm among the neighbor nodes of the suspected node.*

When a node is suspected to be malicious as its reputation value has fallen below a threshold in one or more of its neighbors, the neighbors generate alarms and broadcast them. However, to prevent any possible attempt by a group of colluding nodes to isolate an honest node, a majority voting algorithm is invoked in the neighbourhood of the suspected node. The suspected node is isolated only if the majority of the neighbors agree that the suspected node is to be isolated based on the trust information maintained by them about the suspected node. This algorithm will work fine as along as the majority of the nodes in any neighbourhood are honest (i.e., not compromised).

### 5.2 Trust management scheme

Any currently existing trust and reputation systems can be extended to accommodate the design principles described in Section 5.1. Therefore, details regarding trust metrics and reputation computation is avoided here. Instead, the procedure of building the trust management scheme for the proposed mechanism is discussed in a rather generic way.

*Computation of trust values:* In the proposed system, the trust management module invokes itself automatically after the deployment of the WSN. Each sensor constructs a reputation table corresponding to each of its neighbors and also maintains two counters one for positive and the other for negative experiences for each neighbor with which it has interacted. Initially the trust values for all sensors are assigned to the maximum level as mentioned in Section 5.1.

Since the basic functions of the nodes in a WSN are sensing, processing, reporting and forwarding of data, each sensor is allowed to accumulate its trust value by legitimately participating in these activities. Thus, for example, each successful completion of each of these activities will increase the positive counter for a node by 1, which in turn, will increase the trust level and hence the reputation value of the node. In some applications, the value or significance of the activity is context-based, and thus the increase in the positive counter is activity-dependent. In peer-to-peer networks and mobile ad hoc networks, certain incentives (credit) are given to nodes that answer to the queries on the reputation values of other nodes [15]. However, in sensor networks, as all the sensors belong to the same interest group, responding to the reputation queries is a mandatory responsibility. Accordingly, sensors in the proposed scheme do not increase the positive counters for cooperative participation of the neighbors. On the other hand, the negative counter is increased for each bad encounter that a node experiences with any of its neighbors.

Based on the values of both the positive and negative counters, the reputation value of a sensor is computed and updated accordingly. Use of beta-distribution is very popular in this regard [16, 17]. To defend against more advanced attacks, e.g., *strategic dynamic personality attacks* where a group of malicious node first builds up a good reputation value and then starts misusing it by getting involved in malicious activities till their reputation values just reach the minimum threshold, and again starts building the reputation, more refined trust metrics like TrustGuard [12] can be employed. Invariably there will be trade-off between the computation and communication cost and the accuracy of evaluating the trust values of sensors.

In certain cases, it is possible that the trust value of a sensor is set directly to a value, instead of being computed from the trust metrics. For example, if a node has made violation of the memory consistency rule or in-situ rules (Section 4.3), every node in its neighbourhood will set the its reputation value to 0 and isolate it from the network.

In case of violation of packet delivery rule (Section 4.3), the node that first detects the violation will send an alert message to all the neighbors of the suspected node and the sink node. A higher level intrusion detection action can be taken to identify the source of the attack, because it is usually impossible for a sensor node to detect a malicious neighbor without a multi-layer detection mechanism. This is especially true when the message collision attack (jamming attack) happens below the network layer (i.e., at MAC or physical layer).

*Isolation of malicious sensors:* A very robust and efficient mechanism is employed for isolation of sensor nodes that are identified to be compromised or faulty. The mechanism involves invocation of a majority voting algorithm in the neighborhood of the suspected sensor as follows. Suppose that a sensor detects that one of its neighbor, say *R* is having its reputation value below a pre-determined threshold. The node immediately generates an alert message and broadcasts it to all the neighbors of *R*. The neighbors involve themselves in a voting algorithm where all the neighbors declare their reputation values for *R*. If the majority of the neighbors observe that the reputation value of R is below the threshold then R is isolated from the network and a message to that effect is broadcast in the neighborhood of R. The proposed scheme is thus based on local monitoring of the individual nodes and does not involve computationally complex collaborative monitoring. It has also a very low message overhead.

*On-demand querying trust values of remote sensors:* As the reputation information in the proposed scheme are maintained in the neighborhood, there can be a potential problem when a node needs to communicate to a remote node that is multiple hops away from it. Thus, when a sensor node receives a query from a remote node what should it do? How does it know whether it should communicate with that node or not? In such cases, the receiver node indirectly judges the trustworthiness of the sender on-the-fly. Since each sensor has reputation knowledge about its neighbors only, the reputation of the remote node is computed following a different approach known as '*distance-aware trustworthy route approach*. This approach first estimates the number of hops between the sender (remote) node and the receiver (local) node based on their location information. Then the receiver node tries to find at least one path with the same number of hops to reach the sender node. Each pair of consecutive nodes along that route should have a minimum threshold mutual trust value. If such a route can be successfully found, then the sender node is assumed to be trustworthy and the receiver node starts sending the response to the query.

## 6. Experiments and Results

As the first step of validation of the proposed security mechanism, we have implemented the cryptographic framework for message communication between the sensor nodes and evaluated its performance. Due to hardware constraints of the sensor nodes, we have chosen shared-key cryptography as it is computationally less expensive.

The authentication protocol for the sensor nodes is based on an 8-byte message authentication code (MAC) included in every packet sent by a sensor to the base station. The MAC is computed based on RC5 encryption algorithm. As each sensor has its own shared key with the base station, the base station can verify the authenticity of a message by computing the MAC of the message and comparing it with the MAC on the packet.

The confidentiality in message communication is also achieved by running RC5 in output-feedback mode (OFB). For this purpose, a sensor node uses its secret key and some initialization vector (IV) to calculate a pad. The plaintext is then XORed with the pad to produce the cipher text. OFB is particularly suitable in bandwidth-constrained wireless links since ciphertext is of the same size as the plaintext.

For every packet sent to the base station, the actual payload is encrypted. The combined MAC is computed based on the encrypted payload, the application handler ID, sequence number and source ID. This MAC provides the authentication of the message. The confidentiality and authentication between base station and the sensor nodes establishes a secure communication channel between the sensors and the base station. As a packet is only 30 bytes long, a typical PC (being used as the base station) with Pentium IV, 1024 MB RAM, 2 GHz clock speed, can authenticate 1.27 million packets per second. It can also handle encryption for 1.05 million packets per second. This clearly shows that cryptographic operations will not pose any scalability bottleneck. Instead, key lookup and key set up and storage of expanded keys (which occupies 72 bytes per key) will be. We have simulated the sensor nodes by *SensorSim* extension of network simulator NS2. Each simulated sensor has 8kb of flash memory, 4MHz 8-bit processor and a 900 MHz radio interface. With radio bandwidth of 10 Kbps, each sensor was found to be able to encrypt and authenticate every message it received. In fact the constraining factor is not the computation power but the memory requirement. In fact, the storage of key and buffering of encryption and MAC take 200 bytes out of 512 bytes of available RAM.

Having studied the performance of the cryptographic mechanisms in the sensor network, we plan to take up the performance evaluation of the reputation and trust module of the proposed scheme. For computation of the trust function we plan to consider two extreme cases: (i) when a neighbor node *j* of a node *i* cooperates fully with it and forwards all the packets it receives from i, and (ii) when the node *j* is malicious and drops all the packets that it receives from node *i*. After studying the behaviour of the trust function under these two extreme conditions, we plan to reduce the degree of maliciousness of the node *j* and gradually reduce its packet dropping rate and study the change pattern of the trust function with different rates of packet drop. The convergence of the trust and reputation-related information and associated time for this will be another interesting observation for study. Although the scheme is based on both positive and negative information and thus can be assumed to be immune to false-praise attack and bad-mouthing attack, the resilience of the scheme against such attacks will also be studied.

## 7. Conclusion

Wireless sensor networks are vulnerable to numerous security threats that can endanger their proper functioning. Security support in WSNs is challenging due to limited energy, communication bandwidth, and computational power of the sensor nodes. Given the diversity of WSN applications and possibly different security requirements, a proactive, application-driven approach is needed for making these networks secure. In this paper, we have illustrated that cryptographic solutions are inadequate to tackle the security threats posed by compromised and faulty nodes in WSNs and have proposed a more complete solution based on misbehavior classification, behaviour monitoring and trust management. The proposed scheme is ideally suited for resource-constrained WSNs as it involves minimum computation and communication overhead. The basic cryptographic framework required for this scheme has been implemented on a network simulator and its performance has been evaluated. The results obtained so far have shown the feasibility and effectiveness of the proposed scheme. As a future scope of work, we plan to implement the trust management framework and evaluate its performance. For this purpose, the evaluation parameters are already identified.